\documentclass[12pt]{article}
 \usepackage{graphicx,amsmath,amssymb}
\usepackage{indentfirst}
 \usepackage[usenames]{color}
 \usepackage[colorlinks=true, urlcolor=navyblue, linkcolor=navyblue, citecolor=navyblue]{hyperref}
\usepackage{epstopdf}
\usepackage{appendix}

\definecolor{navyblue}{rgb}{0,0.08,0.45}

\def\Dslash{\raise.15ex\hbox{/}\kern-.7em D}
\def\Pslash{\raise.15ex\hbox{/}\kern-.7em P}

\newcommand{\beq}{\begin{equation}}
\newcommand{\enq}{\end{equation}}
\newcommand{\beqa}{\begin{eqnarray}}
\newcommand{\beqast}{\begin{eqnarray*}}
\newcommand{\enqa}{\end{eqnarray}}
\newcommand{\enqast}{\end{eqnarray*}}
\newcommand{\beml}{\begin{multline}}
\newcommand{\enml}{\end{multline}}

\newcommand{\req}[1]{(\ref{#1})}

\newcommand{\pa}{\partial}

\newcommand{\bec}{\begin{center}}
\newcommand{\enc}{\end{center}}
\newcommand{\beqo}{\begin{quote}}
\newcommand{\enqo}{\end{quote}}

\newcommand{\half}{{\textstyle{\frac{1}{2}}}}

\newcommand{\mbf}[1]{\mathbf{#1}}

\newcommand{\ze}{\zeta}

\newcommand{\la}{\lambda}

\newcommand{\vp}{\varphi}

\newcommand{\De}{\Delta}

\begin{document}

\begin{flushright}
{
\small
SLAC--PUB--15954\\
\date{today}}
\end{flushright}

\vspace{60pt}

\centerline{\Large \bf Light-Front Holography in QCD and Hadronic Physics}

\vspace{20pt}

\centerline{{
Guy F. de T\'eramond,$^{a}$ 
\footnote{E-mail: \href{mailto:gdt@asterix.crnet.cr}{gdt@asterix.crnet.cr}}
Stanley J. Brodsky,$^{b}$ 
\footnote{E-mail: \href{mailto:sjbth@slac.stanford.edu}{sjbth@slac.stanford.edu}}
and
Hans G\"unter Dosch$^{c}$ 
\footnote{E-mail: \href{mailto:gdt@asterix.crnet.cr}{dosch@thphys.uni-heidelberg.de}}
}}

\vspace{30pt}

{\centerline {$^{a}${\it Universidad de Costa Rica, San Jos\'e, Costa Rica}}

\vspace{4pt}

{\centerline {$^{b}${\it SLAC National Accelerator Laboratory, 
Stanford University, Stanford, CA 94309, USA}}

\vspace{4pt}

{\centerline {$^{c}${\it Institut f\"ur Theoretische Physik, Philosophenweg 16, D-6900 Heidelberg, Germany}}

 \vspace{60pt}

\begin{abstract}

We describe  a procedure to extend the light-front holographic approach to hadronic physics to include light-quark masses. The proposed framework allows us to extend the formalism of de Alfaro, Fubini and Furlan to the frame-independent light-front Hamiltonian theory in the approximation where the  dynamics in the invariant transverse variable  is unchanged to first order in the quark masses. The confinement potential follows from an effective theory which encodes the  fundamental conformality of the classical QCD Lagrangian and leads to a semiclassical relativistic light-front wave equation for arbitrary spin. In particular, the  $K$ meson spectrum is successfully described without modifying the emerging confinement scale.  The Wilson loop criteria for confinement is maintained, since for light quark masses a harmonic potential and linear Regge trajectories in the light-front form of dynamics corresponds to a linear potential in the usual instant-form.

\end{abstract}

\newpage

\section{Introduction}

In the  ultrarelativistic limit of zero quark masses one can reduce  the  strongly correlated multi-parton light-front dynamical problem in QCD to an effective  one-dimensional quantum field theory,  which encodes the fundamental conformal symmetry of the classical QCD Lagrangian.  This procedure  leads to a semiclassical relativistic light-front wave equation for arbitrary spin which incorporates essential spectroscopic and non-perturbative dynamical features of hadron physics, similarly to the the Schr\"odinger and Dirac equations in atomic physics~\cite{deTeramond:2008ht,deTeramond:2013it,Brodsky:2013ar}.

A key element in the search for a semiclassical approximation to QCD, in its nonperturbative regime, is the correspondence between the equations of motion in  Anti--de Sitter (AdS) space  and the light-front (LF) Hamiltonian equations of motion for relativistic light hadron bound-states in physical space-time~\cite{deTeramond:2008ht} inspired by the AdS/CFT correspondence~\cite{Maldacena:1997re}. In fact, light-front holographic methods were originally introduced  the matching of the  electromagnetic~\cite{Brodsky:2006uqa} and gravitational~\cite{Brodsky:2008pf} form factors in AdS space~\cite{Polchinski:2002jw,Abidin:2008ku} with the corresponding expressions  derived from LF quantization in physical space time.  This  approach allows us to establish a precise relation between wavefunctions in AdS space and the LF wavefunctions  (LFWFs) describing the internal structure of hadrons.  However the actual form of the effective potential has remained unknown until very recently.

It was been realized~\cite{Brodsky:2013ar} that the form of the effective LF confining potential can be obtained from the framework introduced  by V. de Alfaro, S. Fubini and G. Furlan (dAFF)~\cite{deAlfaro:1976je}, by extending the dAFF formalism  to the frame-independent light-front Hamiltonian theory. It was shown by dAFF  that  a scale can appear in the Hamiltonian while retaining the conformal invariance of the action~\cite{deAlfaro:1976je}. This remarkable result is based on the isomorphism of the algebra of the one-dimensional conformal group $Conf(R^1)$ to the algebra of generators of the group $SO(2,1)$. One of the generators of this group, the rotation in the 2-dimensional space, is compact and has therefore a discrete spectrum with normalizable eigenfunctions. As a result, the form of the evolution operator  is fixed and includes a confining harmonic oscillator potential, and the time variable has a finite range.   As pointed out by dAFF, the relation between the generators of the conformal group and the generators of $SO(2,1)$ suggests that the new scale may play a fundamental role~\cite{deAlfaro:1976je}. In fact, it was shown in Ref.~\cite{Brodsky:2013ar} that there exists a remarkable  threefold connection between the one-dimensional semiclassical approximation to light-front dynamics with gravity in a higher dimensional AdS space, and the constraints imposed by the invariance properties under the full conformal group in one dimension $Conf(R^1)$.  This provides a new insight  into the physics underlying confinement, chiral invariance, and the QCD mass scale.

It was also shown very recently that an effective harmonic potential in the light-front form of dynamics corresponds, for light quark masses, to a linear potential in the usual instant-form~\cite{Trawinski:2014msa}; a result which suggests that the Wilson area law for confinement is also valid for light quarks.  Conversely, for a linear potential in the instant-form, the front-form is a harmonic oscillator, thus the prediction of linear Regge trajectories in the hadron mass square for small quark masses~\cite{Trawinski:2014msa}, in agreement with the observed spectrum for light hadrons.  

Motivated by these recent results, we will discuss in this article how the light-front holographic ideas  can be  extended  in a simple and consistent way to first order in the light-quark masses. In particular, we will show that in this approximation the results are stable; that is,  described with identical values of the gap constant. To first order in the quark masses  the 
dynamics in the invariant transverse variable is unchanged, and the effective LF confining  interaction is given by the effective one-dimensional quantum field theory.

\section{Light-Front Semiclassical Approximation to QCD}

In the font-form of relativistic dynamics~\cite{Dirac:1949cp} the four-momentum generators $P^\mu$ of a hadron $P^\mu = (P^-, P^+,  \mbf{P}_{\! \perp})$, $P^\pm = P^0 \pm P^3$,  are constructed canonically from the QCD Lagrangian by quantizing the system on the light-front at fixed LF time $x^+$, $x^\pm = x^0 \pm x^3$~\cite{Brodsky:1997de}. The LF Hamiltonian $P^-$ generates the LF time evolution 
$P^- \vert \phi \rangle = i \frac{\pa}{\pa x^+} \vert \phi \rangle$,  whereas the LF longitudinal $P^+$ and transverse momentum $\mbf{P}_{\! \perp}$ are kinematical generators.

Each hadronic eigenstate $\vert \psi \rangle$ is expanded in a Fock-state complete basis of non-interacting $n$-particle states $\vert n \rangle$  with an infinite number of components:
$\vert \psi \rangle = \sum_n \psi_n \vert n \rangle$, where the LFWFs $\psi_n$ are boost invariant.  In order to reduce the strongly correlated multi-parton bound-state dynamics to an effective one-dimensional problem it is crucial to identify the key dynamical variable  which controls the bound state~\cite{deTeramond:2008ht},  the invariant mass
of the constituents in each $n$-particle Fock state  $M_n^2 = \left(k_1 + k_2 + \cdots k_n\right)^2$
\begin{equation} \label{M2n}
M_n^2= \sum_{i=1}^n\frac{\mbf{k}_{\perp i}^2 + m_i^2}{x_i},
\end{equation}
where  $\sum_{i=1}^n x_i =1$, and $\sum_{i=1}^n \mbf{k}_{\perp i}=0$.   In fact, the LF wave function is off-shell in $P^-$ and consequently in the invariant mass. Alternatively, it is useful to consider its canonical conjugate invariant variable in impact space. This choice of variable will also allow us  to separate  the dynamics of quark and gluon binding from the kinematics of constituent spin and internal orbital angular momentum~\cite{deTeramond:2008ht}.

For a $q \bar q$ bound state, the invariant mass \req{M2n}, which is also the LF kinetic energy, is $M_{q \bar q}^2 = \frac{\mbf{k}_\perp^2}{x(1-x)}$.
Similarly, in impact space the relevant variable  is  $\zeta^2= x(1-x)\mbf{b}_\perp^2$,  the invariant separation between the quark and  antiquark.
Thus, to first approximation,  LF dynamics  depends only on the boost invariant variable
$M_n$ or $\zeta$,  and the dynamical properties are encoded in the hadronic LF wave function $\phi(\zeta)$
\begin{equation} \label{eq:psiphi}   
\psi(x,\zeta, \varphi) = e^{i L \varphi} X(x) \frac{\phi(\zeta)}{\sqrt{2 \pi \zeta}} , 
\end{equation}
where we have factored out the longitudinal $X(x)$, transverse $\phi(\zeta)$,  and orbital dependence. The normalization of the transverse and longitudinal modes is given by
$ \langle\phi\vert\phi\rangle = \int \! d \zeta  \, \phi(\zeta)^2 = 1$ and $\langle X \vert X \rangle =  \int_0^1 dx \, x^{-1} (1-x)^{-1} X^2(x) =1$.

In the limit of zero quark masses the longitudinal modes decouple  from the  invariant  LF Hamiltonian  equation  $H_{LF} \vert \phi \rangle  =  M^2 \vert \phi \rangle$
with  $H_{LF} = P_\mu P^\mu  =  P^- P^+ -  \mbf{P}_\perp^2$. We obtain the wave equation~\cite{deTeramond:2008ht}
\beq \label{LFWE}
\left(-\frac{d^2}{d\ze^2}
- \frac{1 - 4L^2}{4\ze^2} + U\left(\ze, J\right) \right)
\phi_{n,J,L} = M^2 \phi_{n,J,L},
\enq
a relativistic single-variable  LF Schr\"odinger equation, where $n$ is the number of nodes in $\ze$, $J$ the total angular momentum, and $L$ the internal orbital angular momentum of the constituents. The effective potential $U$ acts on the valence sector of the theory and follows from the systematic expression of the higher Fock components as functionals of the lower ones. This method has the advantage that the Fock space is not truncated, and the symmetries of the Lagrangian are preserved~\cite{Pauli:1998tf}.  The effective interaction potential $U$ is instantaneous in LF time $x^+$, not instantaneous in ordinary time $x^0$, and it represents the complete summation of interactions obtained from the Fock state reduction.

If we compare the invariant mass  in the instant-form in the hadron center-of-mass system, $\mbf{P} = 0$, $M^2_{q \bar q} = 4 \, m_q^2 +  4 \mbf{p}^2$, with the invariant mass in the front-form in the constituent rest frame, $\mbf{k}_q +  \mbf{k}_{\bar q} = 0$ for equal quark-antiquark masses, we obtain the relation found in Ref.~\cite{Trawinski:2014msa} 
\beq \label{pots}
U =  V^2  + 2 \sqrt{\mbf{p}^2 + m_q^2} \,  V +  2 \,  V \sqrt{\mbf{p}^2 + m_{q}^2},
\enq
where we identify $\mbf{p}_\perp^2 = \frac{\mbf{k}_\perp^2}{4 x (1-  x)}$, $p_3 = \frac{m_q (x - 1/2)}{\sqrt{ x(1-x)}}$, and  $V$ is the effective potential in the instant-form. Thus, for small quark masses a  linear instant-form potential $V$ implies a harmonic  front-form potential $U$ and thus linear Regge trajectories. For large  quark masses this relation is still valid for large $q \bar q$ separation, but the non-local mass terms in \req{pots} become important.  One can also show~\cite{Trawinski:2014msa} how the two-dimensional front-form harmonic oscillator potential for massless quarks takes on a three-dimensional form when the quarks have mass  since the third space component is conjugate to $p_3$, which has an infinite range for $m \ne 0$.

\section{Conformal Invariance and Light-Front Hamiltonian Dynamics}

When  extended to light-front holographic QCD~\cite{deTeramond:2008ht,Brodsky:2013ar}, the dAFF framework give important insights into the QCD confining mechanism. It turns out that it is possible to introduce a scale by a redefinition of the quantum mechanical evolution operator while leaving the action  conformally invariant, and consequently to a redefinition of the  corresponding evolution parameter $\tau$, the range of which is finite. Remarkably this procedure determines uniquely the form of the light-front effective potential and correspondingly the  modification of AdS space.  

One starts with the one-dimensional action
\beq
S= \half \int dt \left (\dot Q^2 - \frac{g}{Q^2} \right),
\enq
which  is invariant under conformal transformations in the variable $t$. In addition to the Hamiltonian  $H$  there are two more invariants of motion for this field theory, namely the 
dilatation operator $D$ and $K$, corresponding to the special conformal transformations in $t$.  Specifically, if one introduces the  new variable $\tau$ defined through 
$d\tau= d t/(u+v\,t + w\,t^2)$ it then follows that the the operator $G= u\,H+ v\,D + w\,K$ generates the quantum mechanical evolution in  $\tau$~\cite{deAlfaro:1976je}
\beq
G \vert \psi(\tau) \rangle = i \frac {d}{d \tau} \vert \psi(\tau)\rangle.
\enq
 
In the Schr\"odinger representation~\cite{deAlfaro:1976je}
\beq \label{Htaux} 
G = {1\over 2} u \Big(-{d^2\over dx^2}  + {g\over x^2} \Big) + {i\over 4} v \Big(x {d\over dx} + {d\over dx}x \Big) +{1\over 2}wx^2,
\enq
is the superposition of the `free' Hamiltonian $H$, the generator of dilatations $D$ and the generator of special conformal transformations $K$ in one dimension, the generators of $Conf(R^1)$; namely  $G = u H + v D + w K$.  The conformal group  $Conf(R^1)$ is locally isomorphic to  $SO(2,1)$,  the Lorentz group in 2+1 dimensions. Since the generators of $Conf(R^1)$ have different dimensions, their relations with the generators of SO(2,1) imply a scale, which here plays a fundamental role, as already conjectured in~\cite{deAlfaro:1976je}.

Comparing the dAFF  Hamiltonian \req{Htaux}  with the light-front wave equation \req{LFWE} and identifying the variable $x$ with the light-front invariant variable $\zeta$,  we have to choose $u=2, \; v=0$ and relate the dimensionless constant $g$ to the LF orbital angular momentum, $g=L^2-1/4$,  in order to reproduce the light-front kinematics. Furthermore  $w = 2 \lambda^2$ fixes the confining light-front  potential to a quadratic dependence~\cite{Brodsky:2013ar},  $U \sim \la^2 \, \zeta^2$, and thus from \req{pots} to a linear potential for massless quarks.

\section{AdS Gravity and Light-Front Dynamics}

Anti-de Sitter AdS$_5$ is a five-dimensional  space with negative constant curvature and a 4-dimensional  boundary, Minkowski space-time. In the AdS/CFT correspondence, the consequence of the $SO(2, 4)$ isometry of AdS$_5$ is the conformal invariance of the dual field theory. Recently we have derived wave equations for hadrons with arbitrary spin starting from a  dilaton-modified effective action in  AdS space~\cite{deTeramond:2013it}.    An essential element is the mapping of the higher-dimensional equations  to the LF Hamiltonian equation  found in Ref.~\cite {deTeramond:2008ht}.  This procedure allows a clear distinction between the kinematical and dynamical aspects of the LF holographic approach to hadron physics.  Accordingly, the non-trivial geometry of pure AdS space encodes the kinematics,  and the additional deformations of AdS encode the dynamics, including confinement~\cite{deTeramond:2013it}, and determine the form of the LF effective potential from the precise holographic mapping to light-front physics~\cite{deTeramond:2008ht,deTeramond:2013it}.   The variable $z$ of AdS space is identified with the LF boost-invariant transverse-impact variable $\zeta$~\cite{deTeramond:2008ht,Brodsky:2006uqa,Brodsky:2008pf} thus giving the holographic variable a precise definition in LF QCD. The LF mapping also provides a precise relation between the bound-state amplitudes in  AdS space and the boost-invariant light-front wavefunctions describing the internal structure of hadrons in physical space-time.  One finds  from the dilaton-modified AdS  action the effective LF potential~\cite{deTeramond:2013it,deTeramond:2010ge}
\beq \label{U}
U(\ze, J) = \frac{1}{2}\vp''(\ze) +\frac{1}{4} \vp'(\ze)^2  + \frac{2J - 3}{2 \zeta} \vp'(\ze) ,
\enq
provided that the product of the AdS mass $\mu$ and the  AdS curvature radius $R$ are related to the total and orbital angular momentum, $J$ and  $L$ respectively, according to $(\mu  R)^2 = - (2-J)^2 + L^2$.  The critical value  $J=L=0$  corresponds to the lowest possible stable solution, the ground state of the LF Hamiltonian, in agreement with the AdS stability bound   $(\mu R)^2 \ge - 4$~\cite{Breitenlohner:1982jf}, where $R$ is the AdS radius.  The correspondence between the LF and AdS equations thus determines the LF  confining interaction $U$ in terms of the effective modification of the infrared region of AdS space. The choice of the dilaton profile $\varphi(z) = \lambda z^2$ introduced in~\cite{Karch:2006pv} thus follows  from the requirements of conformal invariance. This specific form for $\varphi(z)$  leads through \req{U} to the effective LF potential 
\beq
U(\ze, J) =   \la^2 \ze^2 + 2 \la (J - 1),
\enq
 and corresponds to a transverse oscillator in the light-front. The term $\la^2 \ze^2$ is determined uniquely by the underlying conformal invariance of classical QCD incorporated in the one-dimensional effective theory, and the constant term  $2 \la (J - 1)$ by the embedding space~\cite{deTeramond:2013it,deTeramond:2010ge}. 
For $\la >0$ , the wave equation   (\ref{LFWE}) has eigenfunctions
\beq \label{phi}
\phi_{n, L}(\zeta) =  \la^{(1+L)/2} \sqrt{\frac{2 n!}{(n\!+\!L\!)!}} \, \zeta^{1/2+L}
e^{- \la  \zeta^2/2} L^L_n(\la \zeta^2) ,
\enq
and eigenvalues
\beq\label{M2SFM} 
M_{n, J, L}^2 = 4 \la \left(n + \frac{J+L}{2} \right),
\enq
an important result also found in Ref.~\cite{Gutsche:2011vb}. This result not only implies linear Regge trajectories, but also a massless pion and the relation between the $\rho$ and $a_1$ mass usually obtained from the Weinberg sum rules~\cite{Weinberg:1967kj}.

\section{Inclusion of Light Quark Masses}

The partonic shift in the hadronic mass from small quark masses follows from the computation of the hadronic matrix element $\langle \psi(P') \vert P_\mu P^\mu \vert \psi({P})\rangle = M^2 \langle \psi(P') \vert \psi({P})\rangle$, expanding the initial and final hadronic states $\vert \psi \rangle$ in terms of their Fock components following the same steps as in Ref.~\cite{deTeramond:2008ht}, but keeping the quark mass  in the kinetic energy terms of the LF Hamiltonian. The result is 
\beq \label{eq:W}
\Delta M^2  =  \left\langle \psi \left\vert \sum_a \frac{m_a^2}{x_a} \right \vert \psi \right \rangle, 
\enq
where $\Delta M^2 = M^2 - M_0^2$ is the hadronic mass shift. Here $M_0^2$ is the value of the hadronic mass computed in the limit of zero quark masses, given by Eq. \req{M2SFM}. This expression is identical to the Weisberger result for a partonic mass shift~\cite{Weisberger:1972hk}. Notice that this result is exact to first order in the light-quark mass if the sum in \req{eq:W} is over all Fock states $n$. 
For simplicity, we consider  the case of a meson bound-state of a quark and an antiquark with longitudinal momentum $x$ and $1-x$ respectively.  To first order in the quark masses
\beq  \label{MKbmq}
\Delta M^2 =  \int_0^1 \! d x \int  \! d^2 \mbf{b}_\perp  
\left( \frac{m_q^2}{x} + \frac{m_{\bar q}^2}{1-x} \right) \left \vert  \psi(x, \mbf{b}_\perp) \right \vert ^2,
\enq
where $m_q$ and $m_{\bar q}$  in \req{MKbmq} are effective  quark masses  from the renormalization due to the reduction of higher Fock states as functionals of the valence state~\cite{Pauli:1998tf},  not ``current" quark masses, {\it i.e.}, the quark masses appearing in the QCD Lagrangian.

The longitudinal factor $X(x)$ in the LFWF \req{eq:psiphi}  can be determined in the limit of massless quarks from the precise mapping of light-front amplitudes for arbitrary momentum transfer $Q^2$. Its form is $X(x) = x^\frac{1}{2} (1-x)^\frac{1}{2}$ \cite{Brodsky:2006uqa}. This expression of the LFWF gives a divergent expression for the partonic mass-shift \req{MKbmq}, and, evidently,  realistic effective two-particle wave functions have to be additionally suppressed at the end-points $x = 0$ and $x = 1$. As pointed out in \cite{Brodsky:2008pg}, the essential dynamical variable which controls the bound state  wave function in momentum space is the invariant mass \req{M2n}. Thus, for the effective two-body bound state the inclusion of  light quark masses amounts to the replacement
\beq \label{Mqbarq}
M^2_{q \bar q} = \frac{\mbf{k}_\perp^2}{x(1-x)} \to \frac{\mbf{k}_\perp^2}{x(1-x)} +\frac{m_q^2}{x} + \frac{m_{\bar q}^2}{1-x},
\enq
in the LFWF in momentum space. This is in fact the correct prescription, since it preserves the invariant properties of the LFWF. In the limit of zero quark masses  the effective LFWF for a  two-parton ground state in impact space is
\beq   \label{LFWFb}
\psi(x, \zeta)  \sim  \sqrt{x(1-x)}~e^{-\half \la \zeta^2},
\enq
where the invariant transverse variable $\zeta^2 = x(1-x) \mbf{b}_\perp^2$ and  $\la > 0$. The longitudinal  factor $\sqrt{x(1-x)}$  is determined from the precise holographic mapping of transition amplitudes in the limit of massless quarks. The  Fourier transform of \req{LFWFb} in momentum-space is
\beq  \label{LFWFk}
\psi(x, \mbf{k}_\perp) \sim  \frac{1}{\sqrt{x(1-x)}}
~e^{- \frac{\mbf{k}_\perp^2}{2 \la x(1-x)}},
\enq
where the explicit dependence of the wavefunction in the LF off shell-energy is evident.

For the effective two-body bound state the inclusion of  light quark masses amounts to the replacement in \req{LFWFk} of the $q \! - \!  \bar q$  invariant  mass \req{Mqbarq}, the key dynamical variable which describes the off energy-shell behavior of the bound state~\cite{Brodsky:2008pg},
\beq  \label{LFWFkm}
\psi(x, \mbf{k}_\perp) \sim  \frac{1}{\sqrt{x(1-x)}}
~e^{- \frac{1}{2 \la} \left( \frac{\mbf{k}_\perp^2}{x(1-x)} + \frac{m_q^2}{x} + \frac{m_{\bar q}^2}{1-x}\right) }.
\enq
Its Fourier transform gives the LFWF in impact space including light-quark masses~\cite{Brodsky:2008pg},
\beq   \label{LFWFbm}
\psi(x, \zeta)  \sim  \sqrt{x(1-x)}~ e^{-  \frac{1}{2 \la} \big(\frac{m_q^2}{x} + \frac{m_{\bar q}^2}{1-x} \big) }  e^{ -\half \la \, \zeta^2},
\enq
which  factorizes neatly  into transverse and longitudinal components. The holographic LFWF \req{LFWFbm} has been successfully used in the description of diffractive vector meson production at HERA~\cite{Forshaw:2012im}, in $B \to \rho \gamma$~\cite{Ahmady:2012dy} and $B \to K^* \gamma$~\cite{Ahmady:2013cva} decays as well as in the prediction of $B \to \rho$~\cite{Ahmady:2013cga} and $B \to K^*$~\cite{Ahmady:2014sva}  form factors.  The LFWF has also been used in Ref.~\cite{Branz:2010ub} to compute the spectrum of  light and heavy mesons.

For excited meson states we can follow the same procedure by replacing the key invariant mass variable in the polynomials in the LFWF using \req{Mqbarq}.  An explicit calculation shows, however,  that the essential modification in the hadronic  mass, from small quark masses, comes from the shift in the exponent of the LFWF. The corrections from the shift in the polynomials accounts for less than 3 \%. This can be understood from the fact that to first order  the transverse dynamics is unchanged, and consequently the transverse LFWF is also unchanged to first order. Thus our expression for the  LFWF 
\beq   \label{LFWFmnl}
\psi_{n, L}(x, \zeta)  \sim  \sqrt{x(1-x)}~e^{-  \frac{1}{2 \la} \big(\frac{m_q^2}{x} + \frac{m_{\bar q}^2}{1-x} \big)}   \zeta^L e^{-\half \la \, \zeta^2}  L^L_n(\la \zeta^2),
\enq
and from \req{MKbmq} the hadronic mass shift $\Delta M^2$ for small quark masses
\beq
\De M_{m_q,m_{\bar q}}^2 =\frac{\int_0^1 dx\,   e^{- \frac{1}{\la}\big(\frac{m_q^2}{ x} +\frac{m_{\bar q}^2 }{1- x}\big)}  \left(\frac{m_q^2}{x} + \frac{m_{\bar q}^2}{1-x}\right)}
{\int_0^1 dx \, e^{- \frac{1}{\la}\big(\frac{m_q^2}{ x} +\frac{m_{\bar q}^2 }{1- x}\big)}} ,
\enq
which is independent of $L$, $S$ and $n$.  For light quark masses, the hadronic mass shift is  the longitudinal  $1/x$ average of the square of the effective quark masses,
 {\it i. e.}, the effective  quark masses from  the reduction of higher Fock states as functionals of the valence state~\cite{Pauli:1998tf}.
The final result for the hadronic spectrum of mesons modified by light quark masses is thus
\beq \label{M2mq} 
M_{n, J, L, m_q, m_{\bar q}}^2 = \De M_{m_q,m_{\bar q}}^2 + 4 \la \left(n + \frac{J+L}{2} \right),
\enq
with identical slope $4 \la$ from the limit of zero quark masses.  In particular, we obtain from \req{M2mq} the spectral prediction for the $J = L+S$ strange meson mass spectrum
\beq \label{M2S}
M^2_{n,L,S} = M^2_{K^\pm} +  4 \la \left(n + L + \frac{S}{2}\right) ,
\enq
where  $M_{K^\pm} \cong 494$ MeV.

As an example,  the  predictions  for the $J = L + S$ light  vector mesons  are  compared with experimental data in Fig. \ref{VMspec}. The data is from PDG~\cite{PDG:2012}.  The spectrum is well reproduced with  identical values  for the mass scale  $\sqrt \la = 0.54$  GeV for the light vector sector.  The model predictions for the $K^*$ sector shown in  Fig. \ref{VMspec} (b) is very good.   However the states $K^*_0(1430)$ and  $K^*_2(1430)$ -- which belong to the  $J = 0$, $J=1$ and $J =2$ triplet for $L=1$, are degenerate. This result  is in contradiction with the spin-orbit coupling predicted by the LF holographic model for mesons; a possible indication of mixing of the $K^*_0$  with states which carry the vacuum quantum numbers.  Fitting the quark masses to the observed masses of the $\pi$ and $K$ we obtain for $\sqrt{\la} = 0.54 ~{\rm MeV}$ the average  effective light quark  mass $m_q= 46$ MeV, $q = u,d$,  and $m_s = 357$ MeV, values between the  current $\overline {MS}$ Lagrangian masses normalized at 4 GeV and typical constituent masses.  With these values one obtains
$\De M^2_{m_q, m_{\bar q}} = 0.067\,  \la , ~ \De M^2_{m_q, m_{\bar s}} = 0.837 \, \la,  ~ \De M^2_{m_s,m_{\bar s}} = 2.062 \, \la$,
 for $\sqrt{\la} = 0.54~ {\rm MeV}$.  
 
\begin{figure}[h]
\centering
\includegraphics[width=7.2cm]{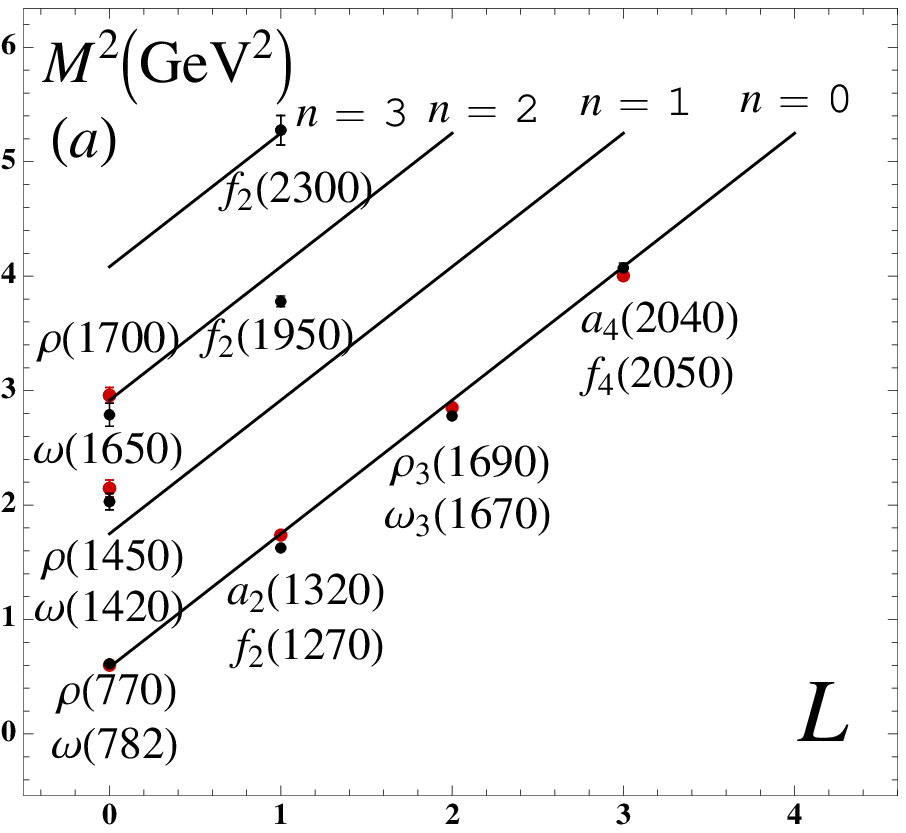}  \hspace{30pt}
\includegraphics[width=7.2cm]{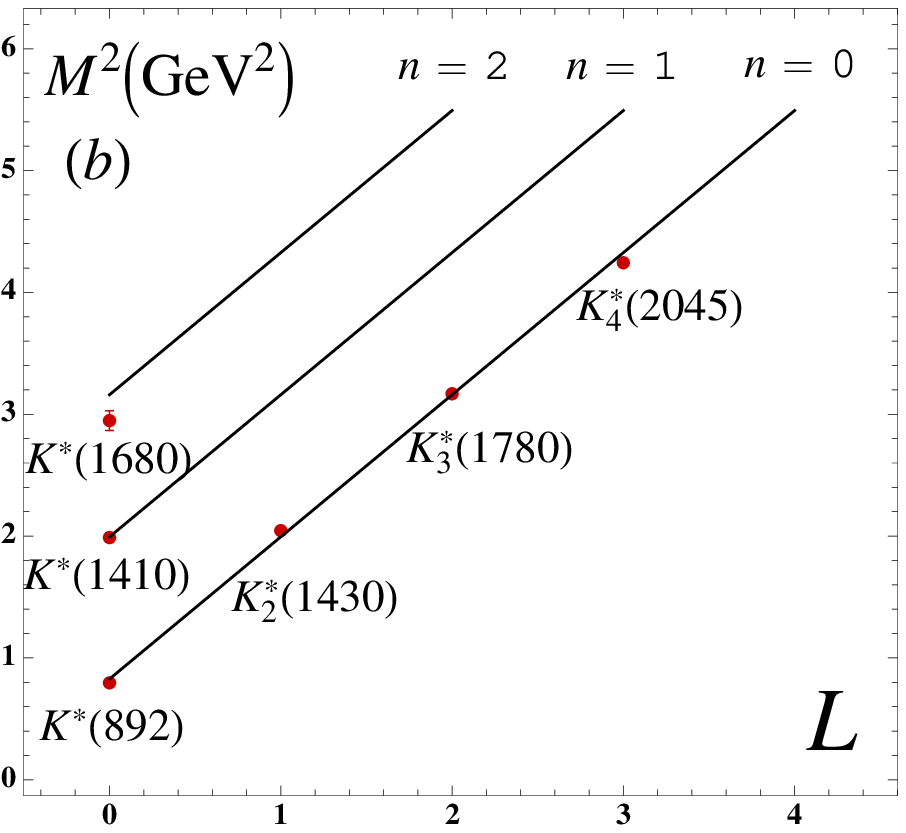}
{ \caption{\small  Orbital and radial excitation spectrum for the light vector mesons:  (a) $I = 0$ and  $I=1$ unflavored mesons and (b)  strange mesons, for $\sqrt \la = 0.54$ GeV.}}
 \label{VMspec}
\end{figure} 

For heavy mesons conformal symmetry is strongly broken and there is no reason to assume that the LF potential in that case is similar to the massless one. Indeed, a simple computation shows that the model predictions for heavy quarks (without introducing additional elements in the model) is not satisfactory.  In fact,  the data for heavy mesons can only be reproduced at the expense of introducing vastly different values for the scale $\la$~\cite{Branz:2010ub,Gutsche:2012ez}. Another important point are the leptonic decay widths. For light quarks the quark masses have little influence on the result, only about 2 \% for the $\pi$ meson and 5 \% for the $K$ meson, but using the formalism also for the $B$ and $D$ mesons leads to widely different values when compared with experiment. For large quark masses the form of the LF confinement potential $U$ cannot be obtained  from the  conformal symmetry of the effective one-dimensional quantum field theory. In this case an important dependence on the heavy quark mass is expected, as  suggested by the relation given by Eq. \req{pots} between the effective potentials in the front-form and instant-form of dynamics.

\section{Conclusions}

The connection of light-front dynamics, its holographic mapping to gravity in a higher dimensional space, and the procedure introduced by de Alfaro, Fubini and Furlan provides new  insights into the physics underlying confinement, chiral invariance, and the origin of the QCD mass scale.  This threefold connection leads to an effective one-dimensional quantum field theory,  which encodes the fundamental conformal symmetry of the  classical QCD Lagrangian.  A mass gap and a confinement scale arise when one extends the formalism  of dAFF to frame-independent light-front Hamiltonian theory, thus leading to emerging confinement.   The resulting light-front potential has a unique form of a harmonic oscillator in the front-form of dynamics and correspond to a linear potential in the usual instant-form.  The result is a relativistic light-front  wave equation for arbitrary spin which incorporates essential spectroscopic and dynamical features of hadron physics.  We have shown how the procedure can be extended to light quarks without modifying to first approximation the transverse dynamics and the universality of the Regge slopes.  As an example we show the new results for the $K^*$ radial and orbital excitations. Recent discussions of light-front holographic predictions are given Refs.~\cite{deTeramond:2014yga,Dosch:2014wxa,Brodsky:2014qqa}.

\section*{Acknowledgements}

Based on an invited talk, presented by GdT at the Rencontres de Moriond, QCD and High Energy Interactions (2014). We thank Joshua Erlich, Stanislaw Glazek and Arkadiusz Trawi\'nski for helpful conversations and collaborations. GdT wants to thank the organizers for the great hospitality at La Thuile.  This research was supported by the Department of Energy, contract DEÐAC02Ð76SF00515.

\end{document}